\documentclass[preprint,aps,tightenlines,showpacs,nofootinbib]{revtex4}
\usepackage{latexsym}
\usepackage[dvips]{graphicx}
\newcommand{\beq}{\begin{eqnarray}}
\newcommand{\eeq}{\end{eqnarray}}
\newcommand{\eq}{eqnarray}

\newcommand{\al}{{\alpha}}
\newcommand{\be}{{\beta}}
\newcommand{\ci}{\cite}
\newcommand{\ga}{{\gamma}}
\newcommand{\Ga}{{\Gamma}}
\newcommand{\ep}{{\epsilon}}

\newcommand{\bepi}{\bar{\epsilon}^{ij}}

\newcommand{\de}{{\delta}}

\newcommand{\la}{{\lambda}}
\newcommand{\La}{{\Lambda}}
\newcommand{\m}{{\mu}}
\newcommand{\n}{{\nu}}

\newcommand{\om}{{\omega}}

\newcommand{\pa}{{\partial}}
\newcommand{\no}{{\nonumber}}
\newcommand{\f}{\frac}
\newcommand{\ra}{\rightarrow}

\newcommand{\we}{\wedge}

\newcommand{\dd}{\de^2 (x-y)}

\begin{document}

\preprint{arXiv:0805.4328v4 [hep-th]}

\title{Constraint Dynamics and
Gravitons in Three Dimensions }

\author{Mu-In Park\footnote{E-mail address: muinpark@yahoo.com}}

\affiliation{ Research Institute of Physics and Chemistry, Chonbuk
National University, Chonju 561-756, Korea }

\begin{abstract}
The complete non-linear three-dimensional Einstein gravity with
gravitational Chern-Simons term
and
cosmological constant
are studied in dreibein formulation. The constraints and their
algebras are computed in an explicit form. From counting the number
of first and second class constraints, the number of dynamical
degrees of freedom, which equals to the number of propagating
graviton modes, is found to be $1$, {\it regardless of} the value of
cosmological constant. I note also that the usual equivalence with
Chern-Simons gauge theory does {\it not} work for general
circumstances.
\end{abstract}


\maketitle

\newpage

\section{Introduction}
In three-dimensional spacetime, there is no propagating ({\it i.e.},
dynamical) degrees of freedom in the bulk with either
Einstein-Hilbert action or gravitational Chern-Simons term (GCS)
with coefficient $1/\mu$. However, it is known that the combined
action with a {\it vanishing} cosmological constant, which is known
as ``topologically massive gravity'' in the literatures, has a {\it
single} propagating, massive, spin-2 mode \ci{Dese:82}. Recently,
there have been several works toward the generalization with a
negative cosmological constant $\La=-1/l^2$ \ci{Li:08,Carl:08}. (For
some earlier related works, see Refs. \ci{Dese:83,Dese:02}.) But the
results do not seem to be in consensus completely. In Ref.
\ci{Li:08}, the wave function for the gravitons and their
corresponding energies are computed for the {\it linearized}
excitations. And it is argued that the theory is
unstable/inconsistent for generic values of $\mu$ due to
``negative'' energies for the massive gravitons. However at the
critical value $\mu l=1$, the massive gravitons ``disappear'' due to
vanishing energies. (See Ref. \ci{Sach:08} for a supporting
analysis.) In Ref. \ci{Carl:08}, the linearized excitations of the
gravitons as well as the scalar and photons are studied in the
``light-front'' coordinates and it is argued that the massive
graviton modes can not be gauged away at the critical value of
$\mu$. (See also Ref. \ci{Grum:08} for a concurrent analysis.)
Rather, at the critical value, it is found that the linearized
topologically massive gravity is equivalent to ``topologically
massive electrodynamics'' with a mass parameter $\mu_{E/M}=2$. And
also, the computations show some splitting of the masses for the
gauge invariant fields
even though there is just one independent degrees of freedom.

On the other hand, it is also known that the three-dimensional
(anti-) de Sitter gravity with or without the GCS term can be
written as a Chern-Simons gauge theory, which does not have the
dynamical degrees of freedom in the bulk \ci{Witt:88}. This seems to
be obviously contradict to the existence of the gravitons in Refs.
\ci{Li:08,Carl:08}.

In this paper, I consider the constraint algebras in the fully
non-linear theory in dreibein formulation. From counting the number
of first and second class constraints, I found that the number of
independent degrees of freedom, which equals to the number of
propagating graviton modes, is $1$, {\it regardless of} the values
of cosmological constant. I do not see any evidence of the
disappearing degrees of freedom at the critical value $ \m l=1$ and
this {\it seems} to support the argument of Ref. \ci{Carl:08}. But,
I note that there is a puzzling feature in this result. I note also
that the usual equivalence with Chern-Simons gauge theory does {\it
not} work for general circumstances.

\section{Hamiltonian formulation}

In this section, I consider the Hamiltonian formulation of the
topologically massive gravity with a cosmological constant
$\La=-1/l^2$, in dreibein formulation
\ci{Dese:82,Dese:83,Dese:02,Witt:88,Krau:06,Solo:06,Park:06}. The
action on a manifold ${\cal M}$, omitting some possible boundary
terms, is given by
\begin{eqnarray}
\label{EHGCS} I=-\frac{1}{16 \pi G} \int_{\cal M} \left[ 2 e^a
\wedge R_a + \f{1}{3 l^2} \ep_{abc} e^a \we e^b \we e^c +\f{1}{\m}
\om^a \we \left(d \om_a +\f{1}{3} \ep_{abc} \om^b \we \om^c \right)
+\la^a \we T_a  \right]
\end{eqnarray}
in form notation with the dreibein and spin-connection 1-{\it forms}
$e^a=e^a_{\m} dx^{\m},~\om^a=\om^a_{\m} dx^{\m}$, respectively
\footnote{The Greek letters ($\mu,\nu, \cdots$) denote the
space-time indices and Latin ($i,j, \cdots$) denote the {\it space}
indices. Latin ($a,b, \cdots$) denote the internal Lorentz indices
and the indices are raised and lowered by the metric
$\eta_{ab}$=diag$(-1,1,1)$ (see Ref. \ci{Park:06} for more details).
I also take the convention $\ep_{012}=-\ep^{012}=1$ and $\ep^{ij}
\equiv \ep^{0ij}$. } The first and the second terms are the
conventional Einstein-Hilbert and the cosmological constant terms,
respectively, with the curvature $R_a=d \om_a + (1/2)\ep_{abc} \om^b
\we \om^c$. The last term is introduced in order to consider the
zero-torsion condition
\begin{\eq}
\label{T=0:comp}
 T_a \equiv d e_a + \ep_{abc}{\omega}^b \wedge e^c=0
\end{\eq}
with 
an auxiliary field $\la^a$ in which $\la^a_i$ is {\it dynamical}
because it multiplies a velocity $\dot{e}_{ai}$ \footnote{It seems
that there are some dual maps between $e^a_{\mu}, ~ \omega^a_{\mu},$
and $ \la^a_{\mu}$ due to the same tensor structure in three
dimensions \ci{Witt:88}. However, due to the difference in the {\it
internal} Lorentz transformation $\de e^a=\ep^{abc} e_b
\theta_c,~\de \omega^a={\cal D} \theta^a$ for the infinitesimal
parameter $\theta^a$, the fields $\la^a$ and $e^a$ would transform
differently also in order that the defining action (\ref{EHGCS}) be
invariant under the transformation. So, the physical contents would
be completely different under the dual maps \ci{Edel:06}.}
\ci{Dese:91}. I have chosen the sign in front of the
Einstein-Hilbert part (with positive Newton's constant $G$) in
agreement with the usual convention in anti- de Sitter space
\ci{Li:08,Dese:83,Dese:02,Witt:88,Krau:06,Solo:06,Park:06,Bana:92,Witt:07}
and all other gravity theories in higher dimensions
\ci{Abbo:82,Brei:82}, but opposite to the original formulation
without cosmological constant \ci{Dese:82,Dese:83} and Ref.
\ci{Carl:08}. The reason for this choice is that its black hole
solution in the $\mu \ra \infty$ limit, {\it i.e.}, Einstein-Hilbert
limit, can be sensible, {\it i.e.}, having``positive'' black hole
mass, only with this sign choice \footnote{In the absence of the
cosmological constant, there is no a priori reason to fix the sign
since there are no gravitons which can mediate the interactions
between massive particles \ci{Dese:84}. The sign is significant only
when the Chern-Simons interaction of (\ref{EHGCS}) is introduced.
The positivity of the gravitational energy \ci{Dese:84} and the
attractiveness of the gravitational interaction \ci{Dese:90} depend
crucially on the overall sign.}

The first-order formulation of the action (\ref{EHGCS}) is given by
\begin{\eq}
I=\int_{\cal M} d^3 x [ \pi^{ai} \dot{e}_{ai} +\Pi^{ai}
\dot{\om}_{ai} + P^{ai} \dot{\la}_{ai} -e^a_0 {\cal H}_a-\om^a_0
{\cal K}_a-\la^a_0 {\cal T}_a-\pa_i \ga^i ]
\end{\eq}
with the conjugate momenta $\pi^{ai},~\Pi^{ai},~P^{ai}$ for
$e_{ai},~ \om_{ai}, ~\la_{ai}$, respectively, and
($\bar{\ep}^{ij}\equiv \ep^{ij}/ 16 \pi G$)
\begin{\eq}
\label{HKT}
  {\cal H}_a &=&  \bar{\ep}^{ij} \left[R_{a ij} +\f{1}{l^{2}} \ep_{abc} e^b_i e^c_j
-2 \ep_{abc}\la^b_i \om^c_j +2 \pa_j \la_{ai} \right], \no \\
  {\cal K}_a &=& \bar{\ep}^{ij} \left[-\f{1}{\m} R_{a ij} +T_{aij}
  -2 \ep_{abc}\la^b_i e^c_j \right],\no \\
 {\cal T}_a &=& -\bar{\ep}^{ij} T_{aij}, \no \\
  \ga^i &=& -\bar{\ep}^{ij} \left[e^a_i \om_{a0}-\f{1}{\m} \om^a_i \om_{a0}-2
\la^a_i e_{a0} \right].
\end{\eq}
The Poisson brackets among the canonical variables are given by
\begin{\eq}
\{ e^a_i (x), \pi^j_b (y) \} =\{ \om^a_i (x), \Pi^j_b (y) \} =\{
\la^a_i (x), P^j_b (y) \} =\de^a_b \de^j_i \de^2 (x-y).
\end{\eq}

\section{Constraint algebras and number of degrees of freedom }

The {\it primary} constraints of the action (\ref{EHGCS}) are given
by
\begin{\eq}
\label{pri}
 &&\Phi^0_a \equiv \pi^0_a \approx 0,~~\Phi^i_a \equiv
\pi^i_a -2 \bepi
\la_{aj} \approx 0, \no \\
&&\Psi^0_a \equiv \Pi^0_a \approx 0,~~\Psi^i_a \equiv \Pi^i_a +
\bepi \left(2 e_{aj}-\f{1}{\m} \om_{aj}\right)
\approx 0, \no \\
&&\Ga^{\m}_a \equiv P^{\m}_a \approx 0,
\end{\eq}
from the canonical definition of conjugate momenta, $\pi^{\m}_a
\equiv \de I/\de \dot{e}^a_{\m},~\Pi^{\m}_a \equiv \de I/\de
\dot{\om}^a_{\m},~P^{\m}_a \equiv \de I/\de \dot{\la}^a_{\m}$. Here,
the weak equality `$\approx$' means that the constraint equations
are used only after working out a Poisson bracket. The conservation
of the constraints $C^0 \equiv (\Phi^0_a, \Psi^0_a, \Ga^0_a)$, {\it
i.e.}, $\dot{C}^0 =\{ C^0, H_C \} \approx 0$, which is a consistency
condition, with the canonical Hamiltonian
\begin{\eq}
H_C = \int d^2 x [e^a_0 {\cal H}_a+\om^a_0 {\cal K}_a+\la^a_0 {\cal
T}_a+\pa_i \ga^i ]
\end{\eq}
produces the {\it secondary} constraints,
\begin{\eq}
\label{sec}
 {\cal H}_a \approx 0, ~~{\cal K}_a \approx 0, ~~{\cal
T}_a \approx 0.
\end{\eq}
With the primary and secondary constraints (\ref{pri}) and
(\ref{sec}), I consider the {\it extended} Hamiltonian which can
accommodate the arbitrariness in the equations of motions due to the
constraints:
\begin{\eq}
\label{H:E}
 H_E =H_C +\int d^2 x [ u^a_{\m} \Phi^{\m}_a + v^a_{\m}
\Psi^{\m}_a + z^a_{\m} P^{\m}_a ].
\end{\eq}
The coefficients $u^a_{\m},~v^a_{\m},~z^a_{\m}$ are determined as
follows, by considering the consistency conditions, $\dot{C}^i =\{
C^i, H_E \} \approx 0$ with $C^i \equiv (\Phi^i_a, \Psi^i_a,
\Ga^i_a)$:
\begin{\eq}
u_{ai}&=&\pa_i e_{0a} -\ep_{abc} (e^b_0 \om^c_i +\om^b_0 e^c_i), \no
\\
v_{ai}&=&\pa_i \om_{0a} -\ep_{abc} \om^b_0 \om^c_i - \m \ep_{abc}
(e^b_0 \la^c_i +\la^b_0 e^c_i), \no
\\
z_{ai}&=&\pa_i \la_{0a} -\ep_{abc} [\la^b_0 (\om^c_i+ \m e^c_i)
+(\om^b_0 +\m e^b_0) \la^c_i] +\f{1}{l^{2}} \ep_{abc} e^b_0 e^c_j.
\end{\eq}
The extended Hamiltonian (\ref{H:E}) reads then as
\begin{\eq}
H_E = \int d^2 x [e^a_0 \bar{\cal H}_a+\om^a_0 \bar{\cal
K}_a+\la^a_0 \bar{\cal T}_a+ u^a_{0} \Phi^0_a + v^a_{0} \Psi^{0}_a +
z^a_{0} P^{0}_a +\pa_i \bar{\ga}^i ]
\end{\eq}
with modified constraints,
\begin{\eq}
\label{sec:mod}
 \bar{\cal H}_a &\equiv& {\cal H}_a -{\cal D}_i \Phi^i_a -\m \ep_{abc}
 \la^b_i \Psi^{ci} + \ep_{abc} \left(- \m \la^b_i +\f{1}{l^{2}} e^b_i  \right) P^{ci} \approx 0, \no \\
 \bar{\cal K}_a  &\equiv& {\cal K}_a -\ep_{abc}
 e^b_i \Phi^{ci} -{\cal D}_i \Psi^i_a -\ep_{abc}  \la^b_i  P^{ci} \approx 0, \no \\
 \bar{\cal T}_a &\equiv& {\cal T}_a -\m \ep_{abc}
 e^b_i \Psi^{ci} -{\cal D}_i P^i_a -\m \ep_{abc}  e^b_i  P^{ci}\approx
 0,
\end{\eq}
and the covariant derivatives ${({\cal D}_i)^c}_a= \de^c_a \pa_i
+{\ep^c}_{ab} \om^b_i $. After a tedious but straightforward
computation I get
\begin{\eq}
\{ \Phi^i_a (x), \Psi^j_b (y) \} &=&2 \bepi \eta_{ab} \dd, \no \\
\{ \Phi^i_a (x), P^j_b (y) \} &=&-2 \bepi \eta_{ab} \dd, \no \\
\{ \Phi^i_a (x), \bar{\cal H}_b (y) \} &=&\f{1}{l^{2}} \ep_{abc} P^{ci}  \dd, \no \\
\{ \Phi^i_a (x), \bar{\cal K}_b (y) \} &=&-\ep_{abc} \Phi^{ci} \dd, \no \\
\{ \Phi^i_a (x), \bar{\cal T}_b (y) \} &=&- \m \ep_{abc} (\Psi^{ci}+P^{ci} ) \dd, \no \\
\{ \Psi^i_a (x), \Psi^j_b (y) \} &=&-\f{2}{\m} \bepi \eta_{ab} \dd, \no \\
\{ \Psi^i_a (x), \bar{\cal H}_b (y) \} &=&- \ep_{abc} \Phi^{ci}  \dd, \no \\
\{ \Psi^i_a (x), \bar{\cal K}_b (y) \} &=&-\ep_{abc} \Psi^{ci}  \dd, \no \\
\{ \Psi^i_a (x), \bar{\cal T}_b (y) \} &=&- \ep_{abc} P^{ci}  \dd, \no \\
\{ P^i_a (x), \bar{\cal H}_b (y) \} &=&- \m \ep_{abc} (\Psi^{ci}  + P^{ci} )\dd, \no \\
\{ P^i_a (x), \bar{\cal K}_b (y) \} &=&-\ep_{abc} P^{ci}  \dd, \no \\
\{ \bar{\cal H}_a (x), \bar{\cal H}_b (y) \} &\approx& \{ {\cal H}_a
(x), \bar{\cal H}_b (y) \} \no \\
&=&\left[ \f{1}{l^{2}} \ep_{abc} {\cal T}^{c} -2 \m \bepi \la_{ai}  \la_{bj} \right] \dd, \no \\
\{ \bar{\cal H}_a (x), \bar{\cal K}_b (y) \} &\approx& \{ {\cal H}_a
(x), \bar{\cal K}_b (y) \} =-\ep_{abc} {\cal H}^{c}  \dd, \no \\
\{ \bar{\cal H}_a (x), \bar{\cal T}_b (y) \} &\approx& \{ {\cal H}_a
(x), \bar{\cal T}_b (y) \} \no \\
&=&[- \m \ep_{abc} ({\cal K}^{c}+{\cal T}^{c} ) +2 \m \bepi \la_{ai}  e_{bj}]  \dd, \no \\
\{ \bar{\cal K}_a (x), \bar{\cal K}_b (y) \}&\approx& \{ {\cal K}_a
(x), \bar{\cal T}_b (y) \} =-\ep_{abc} {\cal K}^{c} \dd, \no \\
\{ \bar{\cal K}_a (x), \bar{\cal T}_b (y) \} &\approx& \{ {\cal K}_a
(x), \bar{\cal T}_b (y) \} =\ep_{abc} {\cal T}^{c}\dd, \no \\
\{ \bar{\cal T}_a (x), \bar{\cal T}_b (y) \} &=&[-2 \m \bepi e_{ai}
 e_{bj}  - \m (e_{ai} P^i_b-e_{bi} P^i_a)] \dd .\
\end{\eq}
Using the above constraint algebras, one can easily see that there
are {\it thirdary} constraints from the consistencies of $\bar{\cal
H}_a \approx 0,~\bar{\cal T}_a \approx 0$ constraints (no additional
constraints from $\bar{\cal K}_a \approx 0$)
\begin{\eq}
\dot{\bar{\cal H}}_a (x) &=&\{\bar{\cal H}_a (x), H_E \}  \no\\
&\approx & -2 \m \bepi \la_{ai} (e^b_0 \la_{bj} -\la^b_0 e_{bj} )
\equiv \Sigma_a \approx 0, \no \\
\dot{\bar{\cal T}}_a (x) &=&\{\bar{\cal T}_a (x), H_E \}  \no\\
&\approx & 2 \m \bepi e_{ai} (e^b_0 \la_{bj} -\la^b_0 e_{bj} )
\equiv \chi_a \approx 0 .
\end{\eq}
The additional constraints $\Sigma_a,~\chi_a$ have the following
non-vanishing brackets:
\begin{\eq}
\{ \Sigma_a (x), \Phi^0_b (y) \}&=&-2 \m \bepi \la_{ai}  \la_{bj}  \dd, \no \\
\{ \Sigma_a (x), \Phi^i_b (y) \}&=&-2 \m \bepi \la_{aj} \la_{b0}  \dd, \no \\
\{ \Sigma_a (x), P^0_b (y) \}&=&2 \m \bepi \la_{ai}  e_{bj}  \dd, \no \\
\{ \Sigma_a (x), P^i_b (y) \}&=&-2 \m \bepi [\eta_{ab} (e^c_0 \la_{cj} -\la^c_0 e_{cj} )
- \la_{aj}  e_{b0} ] \dd, \no \\
\{ \chi_a (x), \Phi^0_b (y) \}&=&2 \m \bepi e_{ai}  e_{bj} \dd, \no \\
\{  \chi_a (x), \Phi^i_b (y) \}&=&2 \m \bepi [\eta_{ab} (e^c_0
\la_{cj} -\la^c_0 e_{cj} )
+ e_{aj}  \la_{b0} ]\dd, \no \\
\{  \chi_a (x), P^0_b (y) \}&=&-2 \m \bepi e_{ai}  e_{bj}  \dd, \no \\
\{  \chi_a (x), P^i_b (y) \}&=&-2 \m \bepi e_{aj}  e_{b0}  \dd .
\end{\eq}
Further investigations of the consistency conditions for the
constraints $\Sigma_a,~\chi_a$, {\it i.e.}, $\{\Sigma_a , H^{'}_E \}
\approx 0,~\{ \chi_a , H^{'}_E \} \approx 0$ with $H^{'}_E =H_E
+\int d^2 x (\al^a \Sigma_a +\be^a \chi_a ) $ do not yield new
constraints but determine the coefficients $u^a_0,~z^a_0,~\al^a,$
and $\be^a$. This completes Dirac's consistency procedure for
finding the complete set of constraints. Although the algebras are
complicated nevertheless one can see that the constraints
$\Psi^0_a,~ \bar{\cal K}_a$ are {\it first} class and the
constraints $\Phi^{\m}_a,~ P^{\m}_a,
~\Psi^i_a,~\Sigma_a,~\chi_a,~\bar{\cal H}_a,~\bar{\cal T}_a $ are
{\it second} class. Here, one might consider some special
configurations of $\la_{a \m}$ and $e_{a \m}$, {\it i.e.}, $A_{ab}
\equiv 2 \bepi \la_{ai}  \la_{bj} \approx 0,~B_{ab} \equiv 2 \bepi
\la_{ai} e_{bj} \approx 0,~C_{ab} \equiv 2 \bepi e_{ai}  e_{bj}
\approx 0$ (neglecting the trivial configurations of
$\la_{ai}=e_{ai}=0$) such that some of these constraints may become
first class or dependent ({\it i.e.}, irregular) \ci{Henn:92}. But,
this is {\it not} relevant to our case (for some related
discussions, see also \ci{Grum:0806,Carl:0807}): $A_{ab} \approx 0$
or $C_{ab} \approx 0$ implies that $\la_{a \m}$ or $e_{a \m}$,
respectively, is not invertible from the fact of $det (\la_{a
\m})=\ep^{abc} \la_{a0} A_{bc} \approx 0$, $det (e_{a \m})=\ep^{abc}
e_{a0} C_{bc} \approx 0$, but the invertibility has been implicitly
assumed from the construction \footnote{In quantum theory, the
non-invertible $e_{a \m}$ might be permitted. See Ref. \ci{Witt:88}
for example.}; moreover, $B_{ab} \approx 0$ would not be generally
true since this implies, from (\ref{HKT}), $\bepi R_{a ij} =0$, {\it
i.e.}, pure Einstein gravity solutions, certainly not a restriction
I wish to consider.

To compute the number of dynamical degrees of freedom I use the
standard formula, at any point $x$,
\begin{\eq}
\label{DOF}
 s=\f{1}{2}(2 n -2 N_1 -N_2),
\end{\eq}
where $2n$ is the number of canonical variables, $N_1$ is the number
of first class constraints, and $N_2$ is the number of second class
constraints. Then, according to the above constraint algebras, I
have $n=9~[e^a_{\m},~\om^a_{\m},~\la^a_{\m}],~N_1=2$, and $N_2=12$
for ``each internal index $a$''. This represents that the system in
terms of the metric $g_{\m \n}=e^a_{\m} e^b_{\n} \eta_{ab}$
\footnote{It would be also interesting to consider the constraint
algebras in the metric formulation directly \ci{Buch:92}, where the
GCS term is a third-derivative order and Ostrogradsky method is
needed \ci{Dese:91}.} has a {\it single} dynamical degrees of
freedom which equals to the number of propagating graviton modes.

Here, in counting the number of degrees of
freedom it would be important to check that they possess a well
defined spectrum. However, unfortunately the kinematic counting of
(\ref{DOF}) gives {\it no} information as to their (in)stabilities.

Before finishing this section, several remarks are in order. First,
the presence of cosmological constant does not modify second class
constraint algebras nor the number of the first class constraints.
So, I do not see any evidence of the disappearing degrees of freedom
at the critical value $\m l=1$: There are two possible scenarios for
this effect, {\it i.e.}, $(a)$ there is an additional ``first''
class constraint, representing a new symmetry, at the critical value
of the cosmological constant \ci{Dese:84b}, $(b)$ there are
compensation terms from the cosmological constant in second class
constraint algebras and the complete compensation is achieved at the
critical value such that the second class constraint becomes first
class \ci{Beng:95}; but, neither of these ``symmetry enhancements''
do not occur in the system. This {\it seems} to support the argument
of Ref. \ci{Carl:08}. Second, the action (\ref{EHGCS}) is not
equivalent to the Chern-Simons gauge gravity \ci{Witt:88,Park:06},
generally. They are equivalent only when one identify
$\la^a_{\m}=e^a_{\m}/ \m l^2$ which changes enormously the
constraint algebras. Actually, the constraint analysis of this
system \ci{Blag:04} leads to $n=6,~N_1=4,~N_2=4$ so that $s=0$, {\it
i.e.}, {\it no} dynamical degrees of freedom. This is consistent
with the fact that the Chern-Simons gauge theory does not have the
dynamical degrees of freedom \footnote{Three-dimensional gravity
without cosmological constant and GCS term can be described by
$ISO(2,1)$ Chern-Simons gauge theory \ci{Witt:88}. However, in the
presence of the GCS term, a second invariant quadratic form for the
Lie algebra is {\it degenerate} such that the gauge theory
formulation does {\it not} exist. This is consistent with the
existence of a graviton in the system \ci{Dese:82}.}. On the other
hand, if one does not introduce the last torsion term in
(\ref{EHGCS}) such that the torsion does not vanish anymore, one has
the same numbers of $N_1=4,~N_2=4$ such that $s=0$ also.

\section{Discussion}

I have shown that counting the number of first and second class
constraints leads to a ``single'' dynamical degrees of freedom for
the {\it metric}, regardless of the value of the cosmological
constant $\La=-1/l^2$. This seems to support the argument of Ref.
\ci{Carl:08} (and Ref.\ci{Grum:08} also), but this is rather
surprising from the following reasons. First, in the context of the
bulk gravity action (\ref{EHGCS}), it is known that there are
critical values of the cosmological constant $| \m l|=1$
\ci{Park:06,Park:06b}, where the characters of the BTZ black hole
solution and matter fields in the black hole background are
dramatically changed. Second, in the context of boundary CFT at the
asymptotic infinity also, the structure of the CFT and its Hilbert
space changes at the critical value. But, at present, there is no
clear understanding of why the fully non-linear constraints do not
show up the above critical features \ci{Park:06}.

\section*{Acknowledgments}

This work was
supported by the Korea Research Foundation Grant funded by Korea
Government(MOEHRD) (KRF-2007-359-C00011).

\newcommand{\J}[4]{#1 {\bf #2} #3 (#4)}
\newcommand{\andJ}[3]{{\bf #1} (#2) #3}
\newcommand{\AP}{Ann. Phys. (N.Y.)}
\newcommand{\MPL}{Mod. Phys. Lett.}
\newcommand{\NP}{Nucl. Phys.}
\newcommand{\PL}{Phys. Lett.}
\newcommand{\PR}{Phys. Rev. D}
\newcommand{\PRL}{Phys. Rev. Lett.}
\newcommand{\PTP}{Prog. Theor. Phys.}
\newcommand{\hep}[1]{ hep-th/{#1}}
\newcommand{\hepp}[1]{ hep-ph/{#1}}
\newcommand{\hepg}[1]{ gr-qc/{#1}}
\newcommand{\bi}{ \bibitem}


\begin{thebibliography}{999}


\bibitem{Dese:82} S. Deser, R. Jackiw, and S. Templeton, Phys. Rev.
Lett. {\bf 48}, 975 (1982);
Ann. Phys. (N.Y.), {\bf
140}, 372 (1982).

\bi{Li:08} W. Li, W. Song, and A. Stromonger, JHEP {\bf 0804}, 082
(2008);
arXiv:0805.3101 [hep-th].

\bi{Carl:08} S. Carlip, S. Deser, A. Waldron, and D. K. Wise,
arXiv:0803.3998 [hep-th].

\bi{Dese:83} S. Deser and J. H. Kay, Phys. Lett. B {\bf  120}, 97
(1983); S. Deser 
 in {\it Quantum Theory of Gravity}, edited by S. M. Christensen (Adam
 Hilga, London, 1984).

\bi{Dese:02} S. Deser and B. Tekin, Class. Quant. Grav. {\bf 19},
L97 (2002);  {\it ibid.} {\bf 20}, L259 (2003).

\bi{Sach:08} I. Sachs and S. N. Soludukhin, arXiv:0806.1788
[hep-th].

\bi{Grum:08} D. Grumiller and N. Johansson, arXiv:0805.2610
[hep-th].

\bi{Witt:88} E. Witten, Nucl. Phys. {\bf B 311}, 46 (1988/89).
\bibitem{Krau:06} P. Kraus and F. Larsen,
JHEP {\bf 0601}, 022 (2006)
.

\bibitem{Solo:06} S. N. Solodukhin,
Phys. Rev. {\bf D 74}, 024015 (2006) 
.


\bi{Park:06} M.-I. Park, Phys. Rev. {\bf D 77}, 026011 (2008);
Phys. Rev. D {\bf 77}, 126012 (2008).

\bibitem{Edel:06}
  J.~D.~Edelstein, M.~Hassaine, R.~Troncoso and J.~Zanelli,
  Phys.\ Lett.\  B {\bf 640}, 278 (2006);
  A.~Giacomini, R.~Troncoso and S.~Willison,
  Class.\ Quant.\ Grav.\  {\bf 24}, 2845 (2007).

\bi{Dese:91} S. Deser and X. Xiang, Phys. Lett. B {\bf  263}, 39
(1991).


\bibitem{Bana:92} M. Banados, C. Teitelboim, and J. Zanelli,
Phys. Rev. Lett. {\bf 69}, 1849 (1992) 
.

\bi{Witt:07} E. Witten, arXiv:0706.3359 [hep-th].

\bi{Abbo:82} L. F. Abbott and S. Deser, Nucl. Phys. {\bf B 195}, 76
(1982); G. W. Gibbons, S. W. Hawking, G. T. Horowitz, and M. J.
Perry, Comm. Math. Phys. {\bf 88}, 295 (1983).

\bi{Brei:82} P. Breitenlohner and D. Z. Freedman, Ann. Phys. {\bf
144}, 249 (1982); Phys. Lett. B {\bf 115}, 197 (1982); L. Mezincescu
and P. K. Townsend, Ann. Phys. {\bf 160}, 406 (1985).

\bibitem{Dese:84} S. Deser, R. Jackiw, and G. 't Hooft, Ann. Phys.
(N.Y.) {\bf 152}, 220 (1984).

\bi{Dese:90} S. Deser, Phys. Rev. Lett. {\bf 64}, 611 (1990).

\bibitem{Henn:92}
  M.~Henneaux and C.~Teitelboim,
  {\it Quantization of Gauge Systems},
(Princeton Univ. Press, Princeton, 1992); J.~A.~Garcia and
J.~M.~Pons,
  Int.\ J.\ Mod.\ Phys.\  A {\bf 13}, 3691 (1998);
O.~Miskovic and J.~Zanelli,
  J.\ Math.\ Phys.\  {\bf 44}, 3876 (2003).

\bibitem{Grum:0806}
  D.~Grumiller, R.~Jackiw and N.~Johansson,
  arXiv:0806.4185 [hep-th].

\bibitem{Carl:0807}
  S.~Carlip,
  arXiv:0807.4152 [hep-th].

\bibitem{Buch:92}
  I.~L.~Buchbinder, S.~L.~Lyahovich and V.~A.~Krychtin,
  Class.\ Quant.\ Grav.\  {\bf 10}, 2083 (1993);
  K.~Hotta, Y.~Hyakutake, T.~Kubota and H.~Tanida,
  arXiv:0805.2005 [hep-th].

\bi{Dese:84b} S. Deser and R. I. Nepomechie, Ann. Phys. {\bf 154},
396 (1984); S. Deser and A. Waldron, Phys. Rev. Lett. {\bf 87},
031601 (2001).

\bi{Beng:95} I. Bengtsson, J. Math. Phys. {\bf 36}, 5805 (1995).

\bi{Blag:04}
M. Blagojevic and B. Cvetkovic, gr-qc/0412134.

\bi{Park:06b} M.-I. Park, Phys. Lett. B {\bf 647}, 472 (2007);
  Phys.\ Lett.\  B {\bf 663}, 259 (2008); Class. Quant. Grav. {\bf
  25}, 095013 (2008).





 \end{thebibliography}
\end{document}